\documentclass[aps,onecolumn,amsmath,amssymb,groupedaddress,notitlepage,floatfix]{revtex4-2}

\usepackage[colorlinks=true, linkcolor=Blue, allcolors=Blue]{hyperref}
\usepackage{graphicx}
\usepackage[T1]{fontenc}
\usepackage[usenames, dvipsnames]{color}
\usepackage{fancyhdr} 
\usepackage{newtxtext,newtxmath}
\usepackage{dcolumn}
\usepackage{bm}
\usepackage{ulem}
\usepackage{subcaption}

\usepackage{amsmath}

\usepackage{array,multirow,makecell}
\setcellgapes{1pt}
\makegapedcells
\newcolumntype{R}[1]{>{\raggedleft\arraybackslash }b{#1}}
\newcolumntype{L}[1]{>{\raggedright\arraybackslash }b{#1}}
\newcolumntype{C}[1]{>{\centering\arraybackslash }b{#1}}

\begin{document}

\title{Propulsive performance of a windsurf-inspired pitching foil}
\author{Gauthier Bertrand}
\author{Tristan Aurégan}
\author{Benjamin Thiria}
\author{Ramiro Godoy-Diana}
\author{Marc Fermigier}
\email{Corresponding author: gauthierbertrand957@gmail.com}
\affiliation{Laboratoire de Physique et M\'ecanique des Milieux H\'et\'erog\`enes (PMMH), CNRS UMR 7636, ESPCI Paris---Universit\'e PSL, Sorbonne Universit\'e, Universit\'e Paris Cit\'e, F-75005 Paris, France}

\begin{abstract}
We study experimentally a symmetrical rigid foil performing pitching oscillations around a mean incidence angle ($\alpha_{m}$) with respect to an incoming flow in a hydrodynamic channel at a constant velocity where the Reynolds number based on the chord of the foil is Re$_{c} = \rho U_{\infty} c / \mu = 14400$. The problem is inspired by the pumping maneuver used by athletes on the new hydrofoil-based windsurf boards. In windsurfing, the pumping mode is used by athletes to produce unsteady propulsion. Because of the complex kinematics and the sail's position relative to the perceived wind, the sail produces thrust or drag and lift. In sailing, the wind perceived by the sail is not parallel to the mean chord. Even if the sail is in a steady mode, it can propel the board by adjusting the mean incidence angle. Few studies investigate the impact of $\alpha_m$ on the mean values of aerodynamic force. The goal of this study is to quantify the forces on this configuration by varying the pitching kinematics characterized by the Strouhal number (St$_{A} = fA/U_{\infty}$, with $f$ and $A$ the frequency and amplitude of pumping, respectively, and $U_{\infty}$ the wind velocity perceived by the sail) from 0 to 0.27, and the mean incidence angle $\alpha_{m}$, from 0$^{\circ}$ to 30$^{\circ}$, of the foil. The force measurements show a high lift production and the delay of the stall angle according to St$_A$ which can be linked to previous studies about the generation of vortices at the leading edge. A general trend of decrease is observed for the drag force coefficient in pitching compared to the static case. For the highest Strouhal numbers tested, the drag coefficient can become negative (thrust) in a range of $\alpha_{m}$ up to 15$^{\circ}$ in specific cases. By using a sport-mimetic approach, we transform the measured lift and drag forces into propulsive and drifting forces, which are the decomposition of the aerodynamic force in the board frame. It is the reference frame in sailing to characterize and optimize physical parameters of the boat such as speed and trajectory. We investigate the impact of unsteady propulsion in upwind conditions. Doing so allows us to investigate race strategies because the generation of propulsion resulting from the drive force is linked to a sideways motion caused by the drift force. We also discuss the impact of using the pumping maneuver rather than the steady propulsion and show several behaviours that could help athletes with decision making.

\end{abstract}
\maketitle
\thispagestyle{fancy}

\section{Introduction}

The competitive practice of sailing and windsurfing has seen a recent revolution with the introduction of new appendixes of hydrofoils that generate lift and keep the board or the boat out of the water for a sufficiently high sailing speed ($U_{\mathrm{boat}} \sim 3$ m/s) \cite{mok2023performance}. This allows one to increase the speed significantly because the wave drag and the hydrodynamic drag are almost suppressed. These innovations include the new iQFOil class introduced for the 2024 Olympic Games.

In order to improve the performance of sailboats and optimize their racing strategy, it is important to characterize their aerodynamic and hydrodynamic response \cite{viola2009force}. Numerous studies on sails have been carried out experimentally in wind tunnels or under real conditions \cite{viola2010full, viola2011sail, fossati2016pressure}. In particular, several works have studied the unsteady effects linked to environmental factors (sea state, wind) or to the dynamic actions of the crew, which can influence the performance of the sails \cite{banks2010measurement, fossati2011experimental, augier2012experimental, augier2013dynamic, Aubin:2016}.  In a laboratory context, a pitching or heaving foil is a reasonably good model to study the generation of unsteady forces which affect the sail propulsion. Young \textit{et al}. \cite{Young:2019} studied experimentally an unsteady propulsion method called sail flicking with a symmetrical rigid foil. With this sailing-mimetic approach, they found a high lift mode in the optimal heaving direction where the foil can produce up to six times the lift in static mode.

\begin{figure}[!t]
    \centering
        \includegraphics[width=1\linewidth]{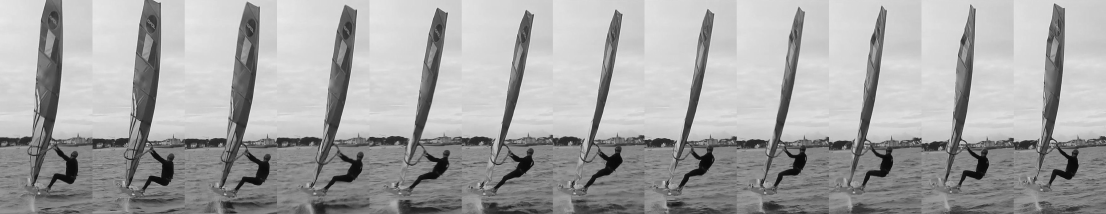}
    \caption{Chronophotography of a one period sequence pumping maneuver of a French athlete during a training session in 2021 at Quiberon. From left to right, we can see the decomposition of the athlete's movement where his center of mass is moving allowing sail's oscillations.}
    \label{fig:pump_seq}
\end{figure}

\begin{figure}[t!]
    \centering
        \includegraphics[width=1\linewidth]{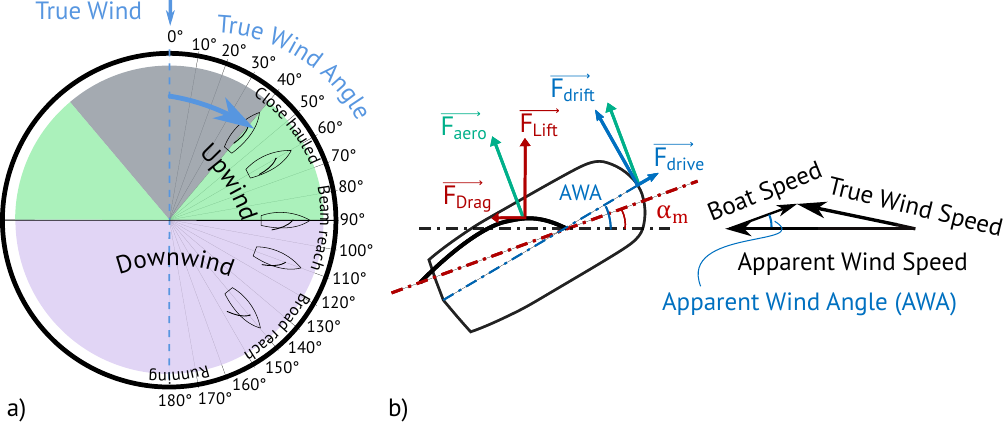}
    \caption{a) The different points of sail of a sailing boat. The angle between the real wind positioned vertically and the longitudinal axis of the boat (reference line) is the true wind angle (TWA). b) Diagram of the dynamics of a sailboat associated with the sailing speed triangle, taking into account the forces applied to the sail and foil. Decomposition of the aerodynamic force (green) in the frame of reference of the boat (blue) and the flow (red). The apparent wind angle (AWA) is the angle between the apparent wind speed (AWS, black dashed line) direction and the boat speed (BS, blue dashed line) direction. $\alpha_{m}$ is the mean angle of attack: the angle between the center line of the sail (red dashed line) and the apparent wind speed (AWS, black dashed line) direction.}
    \label{fig:dyn_boat}
\end{figure}

During race starts or in low wind conditions, particularly after maneuvers like tacking which consist of turning the bow toward and through the wind to go upwind (Figure~\ref{fig:dyn_boat}.a), windsurf athletes employ a technique called pumping to initiate or maintain foiling. This involves rhythmically adjusting the sail's angle relative to the wind through a movement where the athlete moves their center of mass up and down, causing the sail to oscillate and provide intermittent propulsion to keep the board above water (Figure~\ref{fig:pump_seq}). A theoretical and numerical study was conducted to examine the performance of a pumping sail according to complex parameters for windsurfing as a symmetrical foil (NACA 0012) by Zhou \textit{et al}. \cite{zhou2021propulsive}. They analysed the efficiency and the unsteady drive force, which is the projected aerodynamic force in the boat traveling direction (Figure~\ref{fig:dyn_boat}.b), according to flapping parameters and sailing kinematics parameters. In these studies, the unsteady propulsion methods are effective in upwind condition (Figure~\ref{fig:dyn_boat}.a). When windsurf athletes perform the pumping motion, the resulting sail kinematics is three dimensional (3D), but a reasonable leading-order model  can be obtained by limiting the motion to a rotation around the vertical axis. In the present work, we will use such a simplified model: a pitching foil. Using videos from training sessions of the French sailing team we were able to calculate the order of magnitude of the Strouhal number during pumping maneuvers, a non-dimensional number defined as the ratio of the beating speed of the sail over the flow velocity (St$_{A} = fA/U_{\infty}$, $f$ is the pitching frequency and $A$ the beating amplitude). The athletes pump the sail with a frequency $f \sim 1$ Hz and a beating amplitude approximatively equal to the half of the board width $A \sim 0.5$ m. The flow velocity perceived by the sail in the lowest use range is $U \sim 5$ m/s. These physical parameters of the pumping kinematics give St$_A \sim 0.1$. In this study, St$_A$ will be in the range between 0 and 0.27 (Table \ref{tab1}). The orders of magnitude of wind speed, ranging from 5 m/s to 15 m/s, enable the calculation of the Reynolds number associated with the sail chord $c$, Re$_{c} = U_{\infty} c / \nu$, where $U_{\infty}$ is the flow velocity perceived by the sail and $\nu$ is the kinematic viscosity of the fluid. In windsurfing, the range of the Reynolds number is between $\mathrm{10^5}$ and $10^6$ \cite{mok2023performance}, with a chord c of approximately 2 m.

The dynamics of a foil subjected to oscillation has been a matter of interest for several decades for avoiding or reducing undesirable effects such as wing flutter and gust effect, but also to target benefits of the unsteady propulsion \cite{Eldredge:2010}. The study of a flapping foil has  major implications in various fields such as propellers and turbo machinery and wind turbines \cite{mccroskey1982unsteady}, but also in animal locomotion \cite{Triantafyllou2000, izraelevitz2014adding}. In the case of sailing or windsurfing, even without pumping movement, the sail propels the boat, so it is a different problem than the self-propulsion case of animal locomotion. 

A significant amount of experimental works have been conducted to study the physical parameters involved in flapping motion, leading to a better understanding of the generation of unsteady thrust force using flow visualization techniques and force and moment sensors (Platzer \textit{et al}. \cite{Platzer:2008}). The optimal values of thrust force or propulsive efficiency are correlated with inverted Bénard-von K$\acute{\mathrm{a}}$rm$\acute{\mathrm{a}}$n type vortex wakes at the trailing edge \cite{koochesfahani1989vortical, Anderson:1998}. Flow visualization has enabled the identification of different vortex wake structures as a function of flapping amplitude and frequency \cite{zhong2024predicting, godoy2008transitions}. Floryan \textit{et al}. \cite{Floryan:2017} propose scaling laws verified by experiments of thrust, power coefficients and efficiency for a heaving and pitching foil, which summarizes the influence of physical parameters. This is done for a foil with a mean incidence angle of zero, preventing the use of the scaling law of thrust for the windsurf case. The goal of  this paper is to investigate the influence of a wide range of incidence angles in the performance of a pitching foil.

Few experimental studies, such as those by Ohmi \textit{et al}. \cite{ohmi1990vortex, Ohmi:1991}, have examined the influence of a mean incidence angle $\alpha_{m}$ (see Figure~\ref{fig:dyn_boat}.b) around which the foil oscillates for different values of frequency and angular amplitude of the pumping motion. They conducted a study on the vortex wake behind a symmetrical NACA 0012 profile, for two mean incidence angles (15$^\circ$ $\&$ 30$^\circ$) and for each of these mean positions, two amplitudes ($\pm$7$^\circ$ $\&$ $\pm$15$^\circ$) and several frequencies. This provides them with a wide range of Strouhal number values from $0.048$ to $1.03$. This range of St$_{A}$ encompasses that of our study (Table \ref{tab1}). They conclude that the mean incidence angle can contribute more than the position of the pivot or the shape of the foil to the generation of different wakes. Also, they determined that the vortex wake is St$_{A}$ dependent. This study is done without measuring the associated unsteady forces. However, for a motion composed of pitching and heaving other studies discuss the useful effect of adding a non-zero mean incidence angle for maneuvers allowing a strong increase of unsteady side (lift) force coefficient thanks to the effects of the Leading-Edge Vortex (LEV) \cite{schouveiler2005performance}. The generation of LEV affects the propulsion force and efficiency but not the time-averaged lift force \cite{chiereghin2019unsteady}. In addition, the effect of large amplitudes and Reynolds number effect on the generation of forces and wakes for a pure pitching motion have been studied in particular by Zheng \textit{et al}. \cite{Zheng:2021} and Mackowski $\&$ Williamson \cite{mackowski2015direct}, who also highlight the dynamic stall delay. We understand that the effect at the trailing edge is important in the generation of unsteady forces. LEVs appear when a foil is pitching or heaving with large amplitude or when it has a mean incidence angle large enough for this dynamic effect. In certain conditions of foil motion, it is possible to increase the lift force and delay its dynamic stall \cite{Eldredge:2010}. In windsurf, athletes navigate with various condition of wind, from small to high incidence angles of the sail where LEV might be produced, increasing the lift force on the sail.   

We present an experimental study with a symmetrical shape of sail where a pitching movement is applied for various frequencies and amplitudes in a wide range of mean incidence angles to mimic the behaviour of a sail under real sailing conditions, especially upwind conditions (Figure~\ref{fig:dyn_boat}). This study focuses on the coupled effect of the kinematic parameters of pumping (flapping frequency and amplitude) and the mean incidence angle between the mean chord of the sail and the direction of flow, in a context of application to competitive sailing. This sport-inspired unsteady propulsion study covers a parameter range that has not been examined thoroughly.

We define the True Wind Angle (TWA) as in Figure~\ref{fig:dyn_boat}.a to define how the boat is moving according to the True Wind direction, consistent with nautical studies. The points of sail of the boat are also described in Figure~\ref{fig:dyn_boat}.a. A sketch of the balance of forces applied on the sail according to the direction of the wind is shown in Figure~\ref{fig:dyn_boat}.b.

The boat has a driving direction given by the boat speed (BS). The apparent wind speed (AWS) is the composition of the BS and the true wind speed (TWS) and represents the wind perceived by the boat. The apparent wind angle (AWA) is the angle between BS and AWS. The sail produces aerodynamic forces depending on the incidence angle ($\alpha_{m}$). AWA and $\alpha_{m}$ are independent, but both of them influence the drive and the drift forces and so the performance of the boat (Figure~\ref{fig:dyn_boat}.b).

In this study, we will initially examine the impact of pumping amplitude and frequency, as well as the influence of mean incidence, by comparing measurements of lift and drag forces on a symmetrical profile. Subsequently, the forces within the boat's frame of reference (drive and drift forces) are studied in order to identify strategies for implementing pumping in accordance with the prevailing sailing conditions.

\section{Experimental setup and methods}
\begin{figure}[t]
    \centering
        \includegraphics[width=1\linewidth]{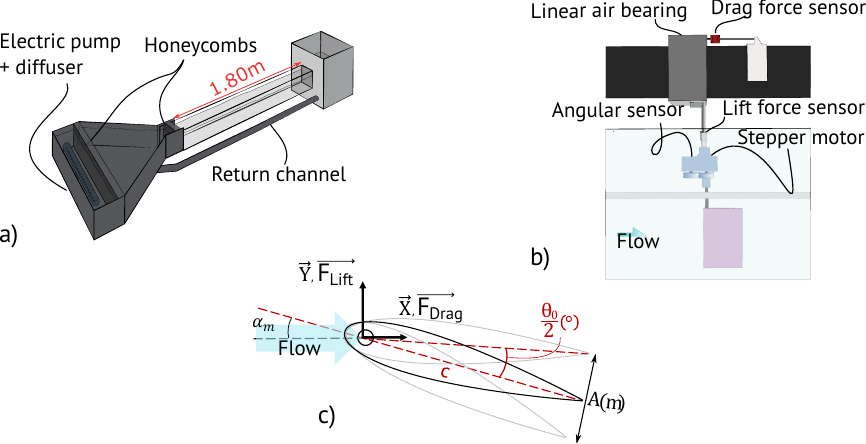}
    \caption{a) Closed loop water channel with a length of $1.80$ m and a cross-section with water of $0.2 \times 0.2$ m$^{2}$ \cite{auregan2023scaling}. b) Acquisition setup to measure forces and kinematics data. c) Sketch in top-view of an experiment. The trailing edge is always more than $5.5$ cm. This value is reached in the extreme case where $\alpha_m =$ 28$^\circ$ and $\theta_0 =$ 11$^\circ$. ($\approx 0.7c$) from the side wall. The top and bottom of the wing are at a distance of 4 cm from the water-air interface and the solid bottom of the channel, respectively.}
    \label{fig:setup}
\end{figure}

Experiments were performed in a free surface water channel in a closed loop with a $0.2$ m by $0.2$ m section. Thanks to the several honeycombs upstream of the test area, the turbulence intensity measured by Particle Image Velocimetry is below $5\%$ (Figure \ref{fig:setup}.a).

The experiments are performed with a NACA 0018 foil 3D printed in PLA with a chord $c = 0.08$ m and a span $s = 0.12$ m, which gives an aspect ratio AR = 1.5. The gap between the top of the foil and the air/water interface is $4$ cm as is the gap between the bottom of the foil and the bottom of the channel. The rotation axis for the pitching movement is located at $0.1$c from the leading edge. The axis is a carbon rod attached to a stepper motor that also drives an angular position sensor. The ensemble rotation motor + foil is mounted on a load sensor (CLZ639HD) that measures the lift component (Y) of the force. This system previously presented is in a sliding connection with an air cushion in relation to the frame (Figure \ref{fig:setup}.b). A second load sensor (FUTEK LSB210) working in traction and compression, located between the linear air bearing and the frame, records the drag force. We sample analogically all the physical parameters and we control the command sent to the stepper motor with a National Instruments card (NI-USB-6221) which allows us to record data at a frequency of 1024 Hz. 

The pitching motion is characterized by the angle between the flow direction and the foil chord as $\theta(t) = \alpha_{m} + (\theta_{0}/2)\sin{(2 \pi f t)}$, where $\alpha_{m}$ is the mean angle of incidence. The angular position sensor measures the mean incidence angle ($\alpha_{m}$) between the foil and the flow, as in a real case with the incidence of the sail and the airflow. The amplitude swept by the trailing edge is $A/2 = (0.9c)\sin{(\theta_{0}/2)}$ (Table \ref{tab1}).

The following experiments were performed at a flow velocity $U_{\infty} = 0.18$ m/s corresponding to a Reynolds number based on the chord Re$_c$ = 14400, (Table \ref{tab1}). By controlling the frequency, amplitude and mean angle of incidence of the sinusoidal motion of the foil, and thanks to the two force sensors, we are able to study unsteady propulsion as a function of the physical and kinematic parameters of the pitching motion described in the Table \ref{tab1}.

\begin{table}[h]
\centering
\small
\begin{tabular}{|c||c||c||c||c|}
\hline
\makecell{Reynolds\\number\\$Re=\rho U_{\infty} c / \mu$} & \makecell{Mean incidence\\angle\\$\alpha_{m}$ (deg)} & \makecell{Reduced\\frequency\\$k = \pi f c / U_{\infty}$} & \makecell{"pitching"\\amplitude\\$A=2(0.9c)\sin(\theta_{0}/2)$ (m)} & \makecell{Strouhal\\number\\$St_{A}=Af/U_{\infty}$}\\
\hline
14400 & $[-8,30]$ & $[2.45,4.20]$ & $[0.0027,0.0162]$ & $[0.045,0.27]$\\ 
\hline
\end{tabular}
\caption{Physical parameters describing our experiments. The flow velocity is fixed at $U_\infty = 0.18$ m/s.}
\label{tab1}
\end{table}

We studied the hydrodynamic forces for different values of $\alpha_{m}$ in the range of $[\![-8 , 30]\!]$, of St$_{A}$ in the range $[0.045 , 0.27]$ and $f$ in the range of $[1.75,3]$ Hz with an increment of 1$^{\circ}$, $0.045$ and $0.25$ Hz respectively (Table \ref{tab1}). For each $\alpha_{m}$, measurements are made by selecting our St$_{A}$ range and the frequency range. As we are testing six frequencies, there will be six experiments per St$_{A}$ value. An amplitude A is coupled to each frequency value tested in order to achieve the desired St$_{A}$ value. Finally, we measure the mean value of lift and drag for the couple (St$_{A}$, $\alpha_{m}$). We run experiments for each case during 30 cycles, and we extract the mean value of forces. 

\begin{figure}[!t]
    \centering
        \includegraphics[width=0.9\linewidth]{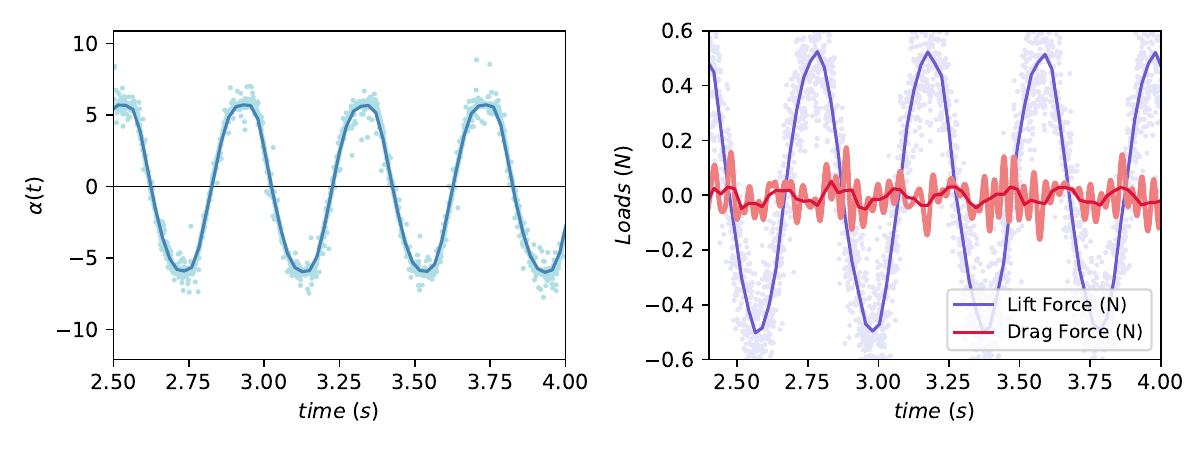}
    \caption{Raw (points) and filtered (lines) data of incidence angle (left) and unsteady forces (right) versus time, for an experiment where $\alpha_{m}$ = 0$^\circ$ , $\theta_{0}$ = 12$^\circ$ and $f$ = 2 Hz ($St_{A}=0.2$). We have smoothed here over $0.1$ s with a Savitsky-Golay filter of order 1. We can observe the phase between the signal of $\alpha(t)$ and the signal of the lift force, which is periodic with $f_{Lift} = f$. The drag force signal is periodic too with $f_{Drag} = 2f$.}
    \label{fig:params}
\end{figure}

We define the aerodynamic coefficients as :
\begin{equation}
    C_{D}=\frac{\overline{F_{Drag}}}{\frac{1}{2} \rho S U_{\infty}^{2}} \ \ \text{and} \ \ C_{L}=\frac{\overline{F_{Lift}}}{\frac{1}{2} \rho S U_{\infty}^{2}}
    \label{aero_coeff}
\end{equation}

$C_{L}$ and $C_{D}$ represent the lift and the drag coefficient, where $\rho$ is the fluid density (1000 kg/m$^{3}$), $S=s \times c$ is the lifting surface (0.0096 m$^{2}$, $U_{\infty}$ is the flow velocity (0.18 m/s) and $F_{Drag}$ and $F_{Lift}$ the drag and the lift force, respectively. Due to the size of the water tank, a correction for the blockage effect is taken into account. For our experimental set-up the blockage area ratio $S_{a}/S_{tank}$ is between $4\%$ and $9\%$, where $S_{tank}$ is the cross flow surface of the water tank ($S_{tank} = 0.04$ m$^{2}$) and $S_a$ is the projected surface orthogonal to the flow of the foil depending on the mean incidence angle. We use in~\eqref{correction} a correction with a quasi-streamlined flow method for  three-dimensional bluff-body changing its mean incidence angle according to the ESDU Technical Committees \cite{esdu:76028}.

\begin{equation}
    \frac{C_{*,c}}{C_{*}} = 1 - \lambda_{1}\lambda_{3}\lambda_{5} \left( 1 + \frac{1}{\lambda_{2}} \frac{s}{c} \right) \frac{c S_a}{S_{tank}^{1.5}} - 0.5C_{D}\frac{S_a}{S_{tank}}    \label{correction}
\end{equation}

$C_{*,c}$ represents the force coefficient corrected and $C_{*}$ the raw coefficient. Where, $\lambda_{1} = 0.72 \times (l/h + h/l)$ is the water tank shape parameter for a three-dimensional flow. $l$ and $h$ are respectively the width and the height of the water tank. $\lambda_{2} \approx 0.83$ is the body shape parameter. $\lambda_{3} \approx \lambda_{3, \ ellipsoid} \times 0.25 + \lambda_{3, \ cone} \times 0.75$ is the body volume parameters. $\lambda_{5} = 1 + 1.1 \times (c/w) \times(\pi/180)^{2} \times \alpha_{m}^{2}$, for taking into account the projected width when $\alpha_{m}$ in degree changes. $w$ is the maximum width of the foil.
\section{Results and discussion}
\subsection{The aerodynamic coefficients on the foil}

\begin{figure}[!t]
    \centering
        \includegraphics[width=0.9\linewidth]{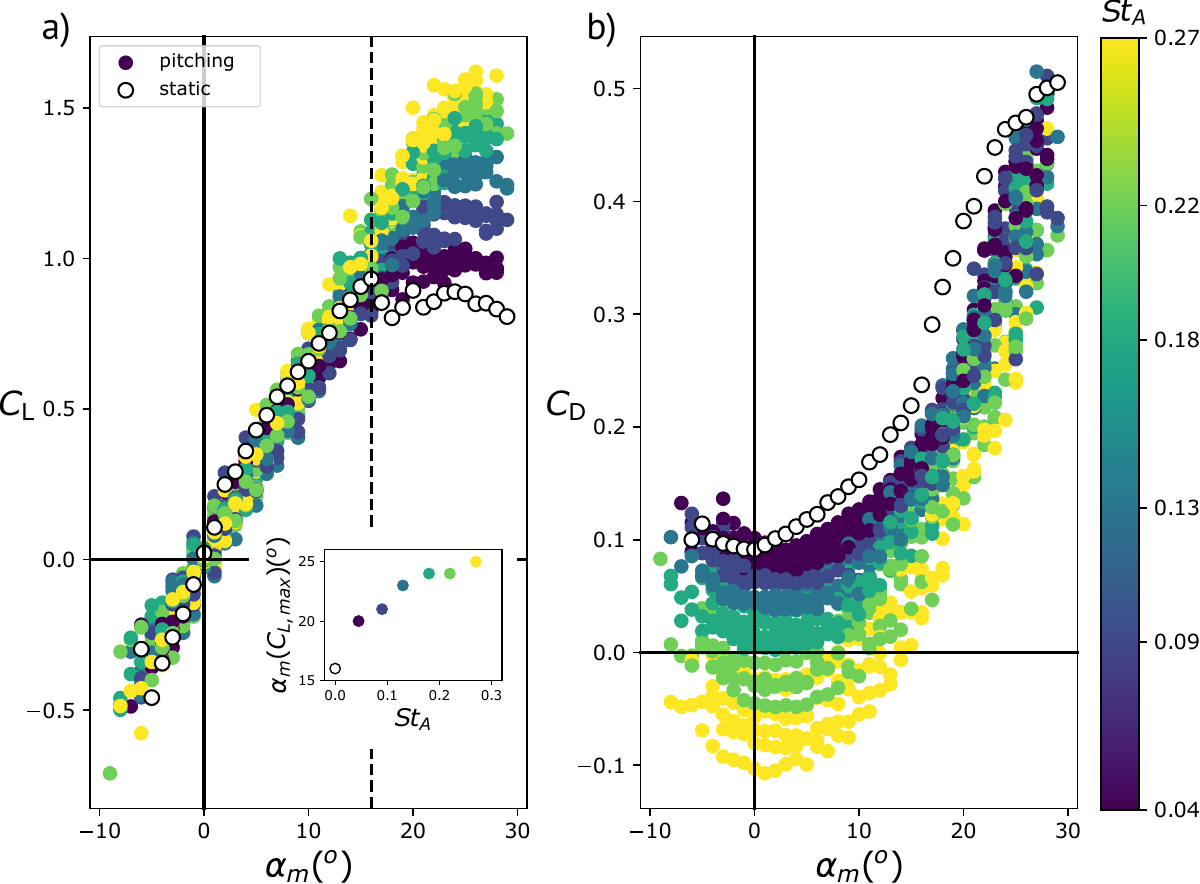}
    \caption{Lift and drag coefficients as a function of the mean incidence angle for a range of Strouhal numbers (colorbar) such as $St_{A} = [0, 0.045, 0.09, 0.13, 0.18, 0.22, 0.27]$ (Table \ref{tab1}). a) Lift coefficients. The vertical dashed line crosses $\alpha_{m} =$ 16$^\circ$ where the static stall appears. In insert, we show the value of $\alpha_{m}$ where the maximum of lift coefficient is reached as a function of St$_{A}$. b) Drag coefficients as a function of the mean incidence angle for the same range of Strouhal numbers.}
    \label{fig:pitch_polar}
\end{figure}

Figure~\ref{fig:pitch_polar} presents the lift and drag coefficients $C_{L}$, $C_{D}$, as a function of the mean incidence angle $\alpha_m$ for all cases tested. The kinematics is represented by the Strouhal number St$_A$ in the third dimension (colorbar). The force coefficients for the static behaviour of the foil are represented by the open circle symbols. The static stall appears at $\alpha_{m} =$ 16$^\circ$, where $\partial C_{L} / \partial \alpha_{m}$ is equal to $0.07$ lower than $\pi^{2}/90$ the theoretical result given by the two-dimensional thin airfoil theory \cite{milne1973theoretical}, which suggests the presence of 3D effects in the flow caused in particular by the small aspect ratio of the foil and the gaps between the top and the water/air interface and the bottom and the solid bottom of the water channel. The collapse of the experimental results for different reduced frequencies $k$ but the same St$_A$, in particular for the lift coefficient in Figure~\ref{fig:pitch_polar}.a, shows that St$_A$ is a more appropriate parameter to describe the effect of pitching on the forces generated than $k$.

The lift coefficient $C_{L}$ increases linearly for all the range of St$_{A}$ at least up to the static stall angle equal to 16$^\circ$. As illustrated in the inset of Figure~\ref{fig:pitch_polar}.a, when St$_{A}$ is increasing, the mean incidence angle associated with the maximum value of lift coefficient $\alpha_{m}(C_{L, \ max})$ increases. Thus, the range where $C_{L}$ grows linearly as a function of $\alpha_{m}$ increases as St$_A$ increases. Even oscillations with the smallest St$_{A}$ tested increase the mean lift produced while delaying the onset of the threshold for high $\alpha_m$. It is worth noting that the first significant effect of the dynamic pitching of the foil is to delay the stall point with respect to $\alpha_{m}$. The most important consequence of this effect is to maintain the growth of $C_{L}$ beyond the stall transition of the static case. This effect is even more pronounced as St$_A$ increases to reach a maximum pitching lift coefficient $(C_{L})_{\mathrm{pitching, \ max}} \approx$ 1.6 (compared to $\approx$ 1 for the static case). Cleaver \textit{et al}. studied the behaviour of the lift force generated by a plunging foil with a non-zero incidence angle. The static stall angle for their foil is 10$^{\circ}$. For an incidence angle between 12.5$^{\circ}$ and 15$^{\circ}$, they presented $C_L$ as a function of St$_A$ and showed a threshold in St$_A$ where $C_L$ does not increase anymore and fataly falls. At an angle of incidence equal to $20^{\circ}$, the drop in $C_L$ as a function of St$_A$ no longer occurs. In the case of plunging, there is therefore a limit to the generation of lift in certain cases \cite{cleaver2009lift, cleaver2012bifurcating}. In our case and so for a different oscillation movement, the drop of lift coefficient as a function of St$_A$ does not appear in the range of St$_A$ tested.

Figure~\ref{fig:pitch_polar}.b shows that for all experiments carried out and for $\alpha_{m} > 0^\circ$, the drag generated when pitching is lower than the static value. More importantly, a drag-thrust transition can be observed: the more we increase St$_{A}$, the more the drag is reduced, until a critical value of Strouhal number $St_A = 0.18$, where drag becomes negative, meaning that the foil generates thrust. At high enough $St_{A} = 0.27$ and for a specific couple of frequency and beating amplitude, the pitching motion generates thrust in a range of $\alpha_{m}$ [$-8^\circ;+15^\circ$]. In the case of a pitching foil with a zero mean incidence angle, maximum propulsive efficiency is reached for $St_A \approx$ 0.3 \cite{Floryan:2017}.

Previous works in the literature have examined the dynamics of oscillating foils at large incidences \cite{ohmi1990vortex, Ohmi:1991}. Seshadri \textit{et al}. \cite{seshadri2023leading} conducted numerical investigations of the flow for the same system of Ohmi \textit{et al}. \cite{ohmi1990vortex, Ohmi:1991}, focusing on one mean incidence angle equal to 30$^\circ$, a beating amplitude equal to $\pm$15$^\circ$, and two reduced frequencies $k= 0.628, \ 3.14$. They demonstrated that as the reduced frequency increases, a LEV is maintained for a longer duration, resulting in a significant enhancement of lift. The amplitudes tested in our study are lower than those observed in the aforementioned studies, but the mechanism for maintaining the leading edge vortex (or vortices) must also be involved in the increase in lift observed in our study. Due to blockage in the channel, the amplitude, denoted by $\theta_{0}/2$, is constrained to a value of at most $11^\circ$. This leads to a maximum incidence angle of $39^\circ$ at a given instant. It would be interesting to ascertain whether the dynamics remain unchanged when the same St$_{A}$ is considered with larger amplitudes. While most of the variation of $C_D$ is captured by plotting it as a function of St$_A$, there is still spread in the data indicating a more complex dependence of $C_D$ with $A^{*} = A/c$ and $k$, especially at high St$_A$. At a given St$_A$, the mean drag increases with the increase of pitching amplitude $A^*$, and therefore decreasing the pitching amplitude $A^*$ and increasing the reduced frequency $k$ is favourable to reduce drag for low mean incidence angles.

Summarizing, Figure \ref{fig:pitch_polar}.a shows that, as St$_{A}$ increases, lift also increases, with the growth being more significant at higher mean incidence angles. This behaviour is observed for all incidence angles, including those beyond the static stall angle. The pitching motion is thus effective in delaying stall. Additionally, as shown in Figure \ref{fig:pitch_polar}.b, increasing St$_{A}$ leads to a decrease in drag, regardless of the amount by which St$_{A}$ is increased. Here we describe the aerodynamic behaviour of a foil subjected to an oscillating movement for the generation of unsteady aerodynamic force by studying the kinematic parameters. In the context of competitive windsurfing, physical and physiological limits constrain the realization of the unsteady propulsion maneuver. Thanks to Figure~\ref{fig:pump_seq}, it is possible to understand that athletes are limited by their body size for the amplitude. During the movement, the athlete must not be destabilized. A too large amplitude could cause the athlete to fall off the board because of the excessive roll moment (around the axis of the sail chord) and poor body position. During the second part of the pumping period, athletes have to bring the sail loaded by the flow back towards them, which limits the flapping frequency according to their physical ability.

\subsection{Generation of propulsion forces in the context of a sailing system}

\begin{figure}[!ht]
    \centering
        \includegraphics[width=0.9\linewidth]{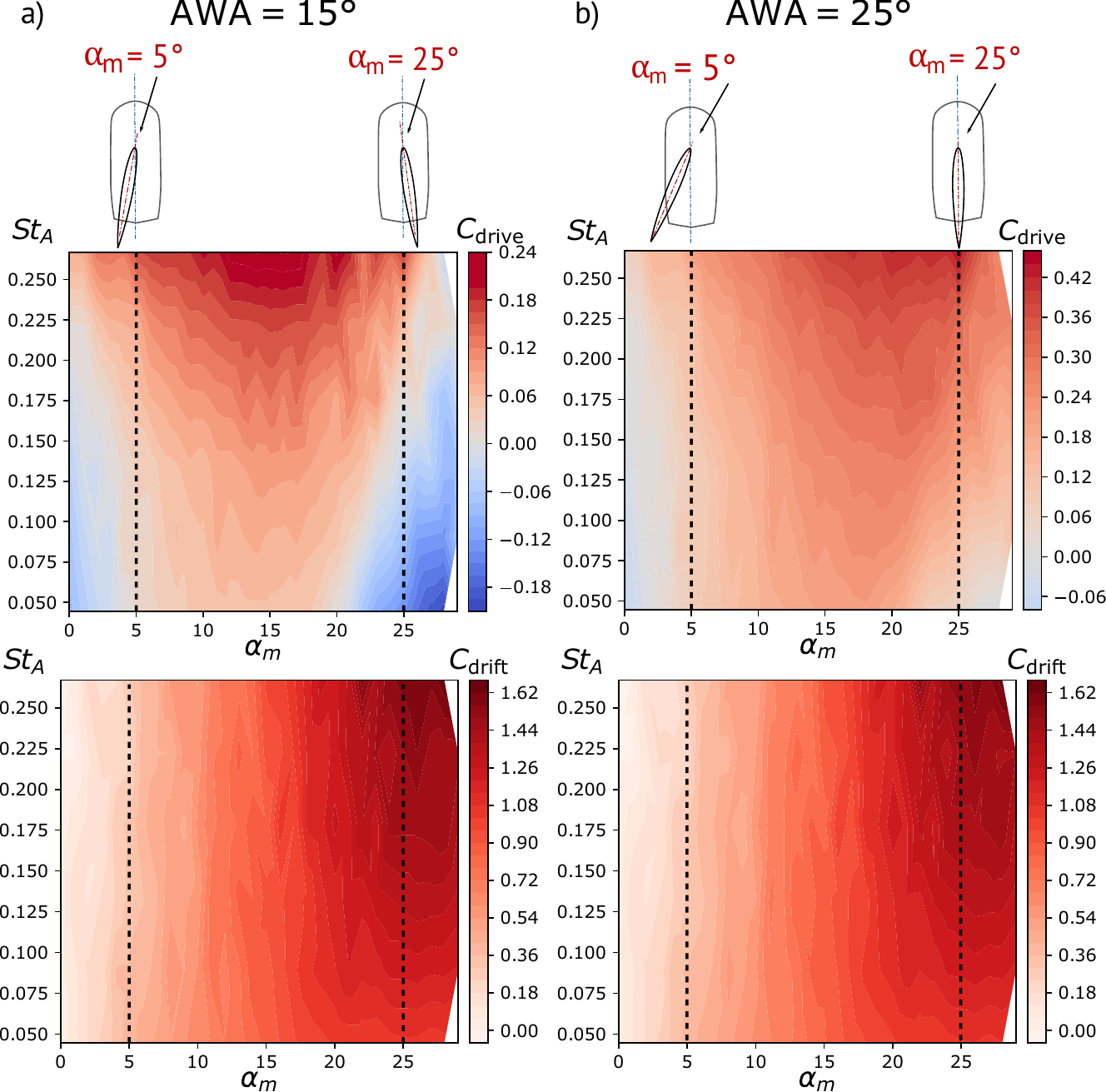}
    \caption{Top: Sketches of four different positions of sailing in upwind conditions. The black arrow illustrates the Apparent Wind Speed (AWS) direction. The red dashed line represents the chord of the sail and the blue dashed line represents the Boat Speed (BS) direction. $\alpha_m$ is the angle between the chord and AWS and the Apparent Wind Angle (AWA) is the angle between AWS and BS. Bottom: Mapping of the sailing coefficients $C_{drive}$ and $C_{drift}$ as a function of pitching parameter St$_A$ and  mean incidence angles $\alpha_m$. The coefficients are averaged by group of St$_{A}$. In upwind conditions for windsurf, AWA is mostly included between $15^\circ$ (a) and $25^\circ$ (b). Note that for $C_{drive}$ (a, b), the color bars do not indicate the same range of coefficients.}
    \label{fig:map_st_alpha}
\end{figure}

In windsurfing, the main performance criterion is the boat speed (Figure~\ref{fig:dyn_boat}). We are therefore interested in the aerodynamic force decomposed in the reference frame of the board with the drive force in the direction of displacement and the drift force perpendicular to the drive force (Figure~\ref{fig:dyn_boat}). These can be written in terms of their respective force coefficients $C_{drive}$ and $C_{drift}$, by dividing the force terms like in ~\eqref{aero_coeff}, as

\begin{align}
    C_{drive}= C_{L}(\alpha_{m}) \sin{(\mathrm{AWA})} - C_{D}(\alpha_{m}) \cos{(\mathrm{AWA})}, \label{drive_coeff}
\end{align}
\begin{align}
    C_{drift}= C_{L}(\alpha_{m}) \cos{(\mathrm{AWA})} + C_{D}(\alpha_{m}) \sin{(\mathrm{AWA})}, \label{drift_coeff}
\end{align}
\noindent where AWA is the Apparent Wind Angle giving the position of the board according to the Apparent Wind direction (see definition in Figure~\ref{fig:dyn_boat}.b). The drive force is directly used for the propulsion and the drift force moves the board sideways. AWA and $\alpha_{m}$ are completely independent, but it is clear that some positions of the sail or the foil according to the direction of the boat are not realistic \cite{Young:2019}. We vary $\alpha_m$ and St$_A$ to examine their effects on the drive and drift force. Figure \ref{fig:map_st_alpha} shows the effective sailing coefficients as a function of the mean incidence angle $\alpha_{m}$ and St$_{A}$ for two values of AWA $15^\circ$ and $25^\circ$, upwind sailing, which are classical positions in windsurfing. The values presented on the maps are averaged according to the Strouhal number for different values of $f$ and $A$.

In both cases (AWA $15^\circ$ and $25^\circ$), Figure~\ref{fig:map_st_alpha} (bottom row) shows that for $\alpha_{m} \lessapprox 20^{\circ}$   $C_{drift}$ is more sensitive to an increase of the mean incidence angle than to an increase of the Strouhal number. For $\alpha_{m} \gtrapprox 20^{\circ}$, $C_{drift}$ depends mainly on St$_{A}$. With the goal of maximizing the sailing speed and reducing the racing time, both $C_{drive}$ and $C_{drift}$ need to be optimized. These will be the tools to manage the boat's course, depending on the race strategy. For the two studied AWA values representing the boundaries of the typical upwind navigation range, the $C_{drift}$ values are almost identical. This is due to the fact that $C_{L}\cos(\mathrm{AWA}) \gg C_D \sin(\mathrm{AWA})$ because $\sin{(\mathrm{AWA})} < \cos{(\mathrm{AWA})}$ and $\cos{(\mathrm{AWA})} \approx 1$.
Figure~\ref{fig:map_st_alpha}.a shows that pitching generates drive force coefficient with a maximum $C_{drive, \ max} \approx 0.24$. When AWA is $15^{\circ}$, it becomes evident that as St$_A$ increases, the range of $\alpha_m$ capable of generating propulsion ($C_{drive} > 0$) expands. For instance, at $St_A = 0.05$, the propulsion range is $5^{\circ} \leq \alpha_{m} \leq 20^{\circ}$ with the lowest values of $C_{drive}$ reached compared with those reached for the other Strouhal numbers of the study. At $St_A = 0.25$, the range broadens to $0^{\circ} \leq \alpha_{m} \leq 25^{\circ}$. 
As AWA increases, the range of $\alpha_{m}$ where $C_{drive} > 0$ increases too, as can be seen in Figure~\ref{fig:map_st_alpha}, comparing the top row of panel (a) (AWA =  $15^{\circ}$) and panel (b) ($25^{\circ}$). Nonefficient zones where the propulsion generated by pumping is less than or close to zero ($C_{drive} \leq 0$) are located at low $\alpha_{m}$ close to zero and at the largest angles studied here. The negative zone for the largest angles comes from the influence of the term $C_{D}(\alpha_{m}) \cos{(\mathrm{AWA})}$ in Equation~\ref{drive_coeff}. For AWA $= 15^{\circ}$ in the low $\alpha_m$ range (top left panel), the increase of the driving force with St$_A$ is very visible and corresponds to the drag to thrust transition mentioned on Figure~\ref{fig:pitch_polar}. Moving from AWA $= 15^{\circ}$ to AWA $= 25^{\circ}$, the term $C_{D}(\alpha_{m}) \cos{(\mathrm{AWA})}$ decreases, and it becomes easier to generate drive force.

\begin{figure}[!t]
    \centering
        \includegraphics[width=1\linewidth]{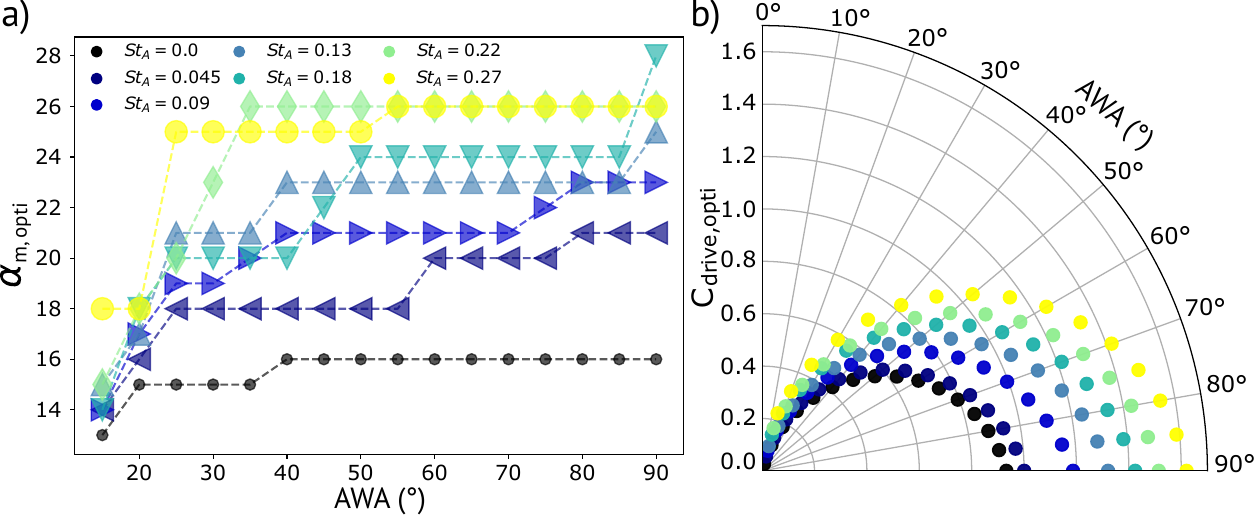}
    \caption{a) $\alpha_{\mathrm{m, \ opti}}$ as a function of AWA. b) Polar of the optimum drive coefficient $C_{\mathrm{drive, \ opti}}$ as a function of AWA. They are represented for each value of St$_{A}$ experimentally studied. For each AWA we only plot the optimal value based on the data presented in Figure~\ref{fig:pitch_polar}, and we determine for each St$_{A}$ the mean incidence angle $\alpha_{\mathrm{m, \ opti}}$ according to $C_{\mathrm{drive, \ opti}}$.}
    \label{fig:force_coeff_opti}
\end{figure}

To optimize the traveled speed, we want to maximize the $C_{drive}$ coefficient. From the working map (St$_A$, $\alpha_{m}$) of $C_{drive}$, we now determine the optimal drive coefficient $C_{\mathrm{drive, \ opti}}$ for each value of St$_{A}$ and the mean incidence angle $\alpha_{\mathrm{m, \ opti}}$ associated with this optimal value, which are illustrated in Figure \ref{fig:force_coeff_opti}. For each value of St$_{A}$, the experimental data shows that $\alpha_{\mathrm{m, \ opti}}$ increases as a function of AWA. In Figure~\ref{fig:force_coeff_opti}.b, the driving force is always larger when pitching compared to the static case (black, $St_{A} = 0$).

Let us look at the case where $C_{drive} \approx 0.6$ in Figure~\ref{fig:force_coeff_opti}.b. Increasing St$_A$ will make it possible to maintain the value of $C_{\mathrm{drive, \ opti}}$ while reducing AWA, in this case potentially going from AWA $= 55^{\circ}$ in static to AWA $= 35^{\circ}$ if the oscillation rises to a St$_A$ of 0.3. This corresponds to a variation in the angle of incidence, $\alpha_{\mathrm{m, \ opti}}$, between $16^{\circ}$ and $25^{\circ}$. Using unsteady propulsion, the boat can maintain its forward momentum while sailing upwind more efficiently. Reducing the AWA reduces the distance traveled and potentially reduces the number of tack changes needed. Tacking implies in most cases that the board is not lifted anymore, so reducing the distance traveled and potentially the number of tack changes can definitely reduce the race time.

Another race strategy can be highlighted using Figure~\ref{fig:force_coeff_opti}.a. Athletes may want to maintain their trajectory and therefore maintain AWA constant while increasing $C_{\mathrm{drive, \ opti}}$. Let's take the case where they want to keep an AWA $= 25^{\circ}$. The athletes will then have to modify the mean incidence angle according to how they increase the St$_A$ associated with an increase in $C_{drive}$. This strategy will then result in a threefold increase of the driving force coefficient from 0.18 up to at most 0.44 for as long as the pumping motion is maintained.

In this section, we have discussed how to generate the most beneficial pitching possible, as well as various navigation strategies that can optimize the trajectory and propulsion of the craft. We now need to compare unsteady and steady propulsion. 

\begin{figure}[!ht]
    \centering
        \includegraphics[width=0.9\linewidth]{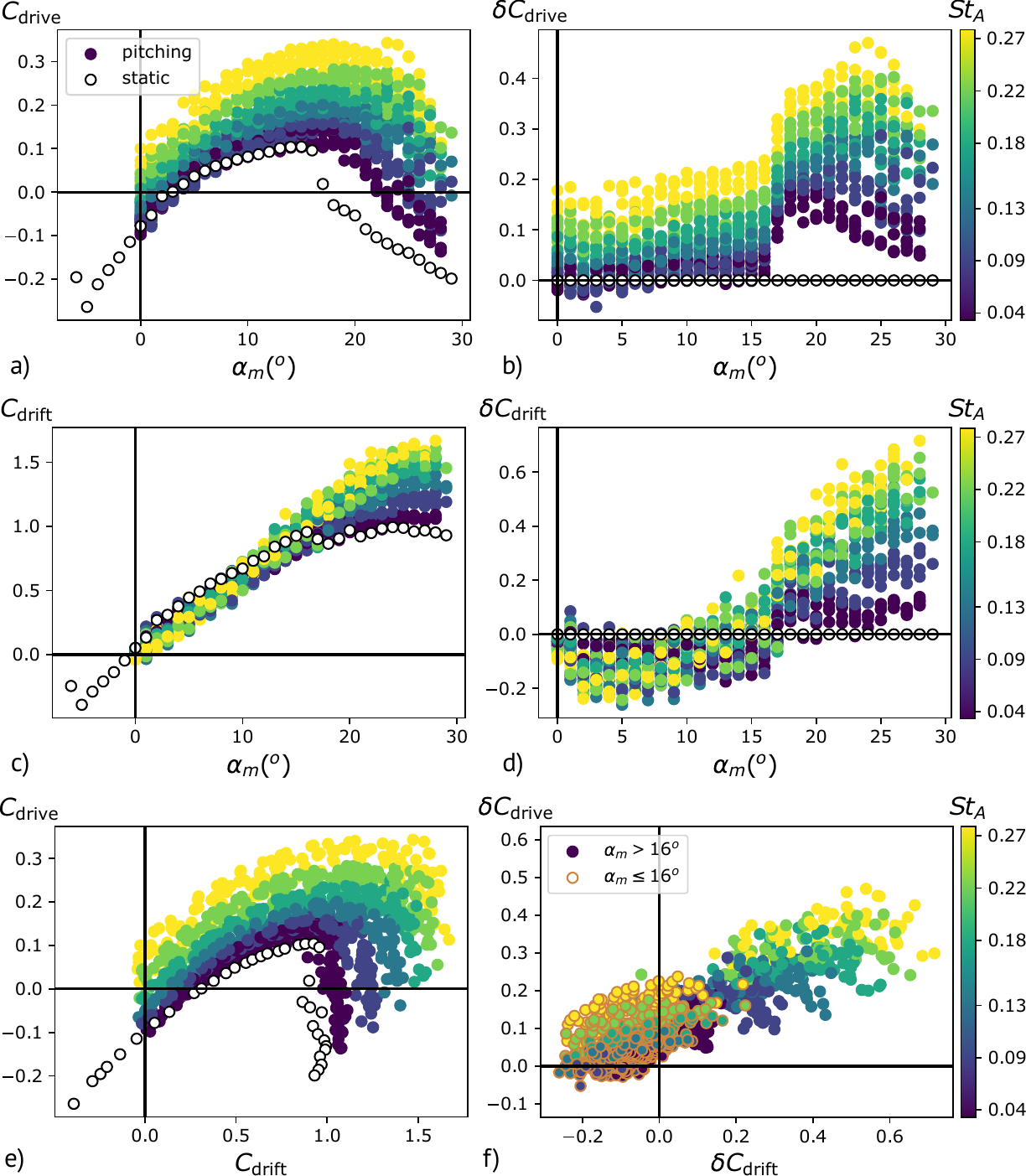}
    \caption{Sailing force polars of drive and drift coefficients and impact coefficient polars as a function of the mean incidence angle for $AWA=20^{\circ}$. a) $C_{drive}$, b) $\delta C_{drive}$, c) $C_{drift}$ d) $\delta C_{drift}$.}
    \label{fig:polar_cx_vs_cy}
\end{figure}

We present in Figure~\ref{fig:polar_cx_vs_cy}.a and .c the sailing force polars as a function of the mean incidence angle for AWA $=20^{\circ}$, which is a classical position of upwind in windfoil. These polars summarize the behaviour of the driving and drifting forces. In order to compare the effect of pitching propulsion with that of stationary propulsion, we introduce a parameter that quantifies the impact of pitching compared with standard propulsion, defined as:

\begin{equation}
    \delta C_{*} = C_{*, pitch} - C_{*, static}. 
    \label{delta_difference}
\end{equation}

We plot the sailing force polars of the impact coefficients of drive $\delta C_{drive}$ and of drift $\delta C_{drift}$ as a function of $\alpha_{m}$ in Figure~\ref{fig:polar_cx_vs_cy}.b and .d.

In Figure~\ref{fig:polar_cx_vs_cy}.a, the static polar of $C_{drive}$ reveals a stall occurring at $\alpha_{m} =$ 16$^{\circ}$, beyond which no propulsion is generated ($C_{drive} \leq$ 0). When athletes employ pumping techniques, they notably enhance the drive force compared to the static condition, with particularly significant improvements at mean incidence angles exceeding 16$^{\circ}$.
In the lowest range of $\alpha_{m}$ some pitching cases seem to be ineffective compared to the steady propulsion. With the use of the impact coefficient,  Figure~\ref{fig:polar_cx_vs_cy}.b highlights these ineffective cases. We can see for $\alpha_{m} \leq$ 8$^{\circ}$ some impact coefficients are negative. It shows an inefficiency to pitch for these cases, for example at $\alpha_{m} =$ 2$^{\circ}$ (Figure~\ref{fig:polar_cx_vs_cy}.c) where St$_A =$ 0.045 or St$_A =$ 0.09. For $8^{\circ} < \alpha_{m} \leq 16^{\circ}$, $\delta C_{drive}$ depends weakly on $\alpha_{m}$ but depends on St$_A$. Figure~\ref{fig:polar_cx_vs_cy}.c shows the sailing polar of $C_{drift}$ as a function of $\alpha_{m}$. $C_{drift}$ increases almost linearly until $\alpha_{m} \approx$ 16$^{\circ}$. As St$_A$ increases, so does the range of $\alpha_m$ where $C_{drift}$ increases linearly. As can be seen in the Figure~\ref{fig:polar_cx_vs_cy}.d, $\delta C_{drift}$ becomes positive for all St$_{A}$ at $\alpha_{m} =$ 18$^{\circ}$. Below this value of $\alpha_{m}$, using unsteady propulsion reduces the drift coefficient which is a major information for the athletes' positioning strategy on the water.

Finally, we discuss this impact coefficient applied to the drive and drift forces at AWA $=20^{\circ}$. Figure~\ref{fig:polar_cx_vs_cy}.e shows the operating polar of $C_{drive}$ versus $C_{drift}$, while that of $\delta C_{drive}$ as a function of $\delta C_{drift}$ is shown in Figure~\ref{fig:polar_cx_vs_cy}.f, summarizing the impact of pitching on propulsion. We separate the values into two parts: those for $\alpha_{m} \leq 16^\circ$ (circled in orange) and those for $\alpha_{m} > 16^\circ$.

In Figure~\ref{fig:polar_cx_vs_cy}.b, for $\alpha_{m} \leq 16^\circ$ a few values of $\delta C_{drive}$ are negative and so in these cases pumping the foil is detrimental. However, within this range of $\alpha_{m}$ and during pitching, for most of the studied cases, it is possible to reduce the drift force generated by the steady propulsion of the sail, a capability applicable in most scenarios. During a race, sail pumping can thus enable athletes to maintain their course more effectively by reducing $C_{drift}$ without compromising the propulsive force $C_{drive}$, ultimately resulting in shorter race times.
In the second part at higher $\alpha_{m} > 16^\circ$, a distinct behaviour emerges with respect to St$_A$, where increasing St$_A$ maximizes $\delta C_{drive}$ and maximizes $\delta C_{drift}$ too.

In the case where AWA = $20^{\circ}$ (Figure~\ref{fig:polar_cx_vs_cy}), let us take a position where the chord of the sail and the direction of travel of the boat are aligned. This gives a $\alpha_m = 20^{\circ}$. From Figure~\ref{fig:polar_cx_vs_cy}.b and .d and considering a moderate range 0.04 < St$_A$ < 0.13, we can quantify in order of magnitude the impact of pumping on the drive force coefficient. It would then be possible to obtain $\delta C_{drive}$ between 0.1 and 0.25 (Figure~\ref{fig:polar_cx_vs_cy}.b and .d), giving an increase in drive force by pumping $\delta F_{drive} = 1/2 \rho U_{\infty}^{2} S\delta C_{drive}$ between 40 N and 100 N, for U $\approx 10$ m/s, S = 8 m$^{2}$ and $\rho_{air}$ = 1.02.

\newpage
\section{Conclusion}
We experimentally investigated the dynamic response of a foil oscillating about its vertical axis, examining how the mean angle of incidence impacts transient forces. Within the studied physical parameter ranges (Table~\ref{tab1}), we discovered the potential to generate lift coefficients nearly double the static values at mean angles of incidence exceeding the static stall angle, with a pronounced dependence on St$_{A}$. Moreover, pitching the foil delays the stall angle and increases $C_{L}$ past the static stall value. Up to the static stall angle, the increase in $C_{L}$ is not primarily governed by St$_{A}$. However, for low $\alpha_{m}$ ranges, we observed a classical result where $C_{D}$ depends on St$_{A}$ and exhibits a drag-propulsion transition. The relationship is not singularly determined by frequency, amplitude, or Strouhal number, but rather by their combined interaction, as highlighted by Floryan \textit{et al}. \cite{Floryan:2017}.
Drawing an analogy with sailing, particularly windsurfing, we introduced coefficients $C_{drive}, \ C_{drift}$ to characterize the forces applied in the boat's frame. We found that $C_{drift}$ depends on $\alpha_{m}$ rather than St$_{A}$, while $C_{drive}$ demonstrates an increasingly expansive operating range ($C_{drive}>0$) that grows with St$_{A}$.
From these observations, we determined the maximal propulsive force and its associated optimal mean incidence angle ($\alpha_{\mathrm{m, \ opti}}$) for upwind conditions (AWA $<90^{o}$). Critically, studying drive force in conjunction with drift force becomes essential for racing strategic considerations. The potential scenarios include either changing position by increasing the drift force or increasing speed without altering course by focusing solely on forward force.
The final segment of our study presents the coefficient differences between pitching and static conditions, highlighting the performance gains achieved through unsteady propulsion. This study was conducted for a symmetrical foil to mimic the sail. It would be very interesting to investigate the impact of $\alpha_m$ and St$_A$ on unsteady propulsion for more complex shapes of foil or sail, (for instance asymmetrical or 3D foils) using a load sensor and flow visualization to compare them. The flexibility of the foil could be taken into account as well. Using a torque sensor on the pitching axis would allow to study the power consumption and the efficiency, which can be used for maintaining over time the pumping maneuver and be linked with the physiological demands.

\section*{Acknowledgments}

This work was partially supported by the French National Research Agency with grant ``Sport de Très Haute Performance'' ANR 19-STHP-0002. The authors would like to thank Amaury Fourgeau for technical support. We are grateful to Benoît Augier and Baptiste Lafoux for fruitful discussions and scientific suggestions.

\bibliography{biblio_pitch.bib}
\end{document}